\documentclass{IOS-Book-Article}
\usepackage{mathptmx}
\usepackage{amsmath} 
\usepackage{algorithm} 
\usepackage{algpseudocode} 
\usepackage{quantikz}
\usepackage{soul}\setuldepth{article}
\usepackage{fancyhdr}

\fancypagestyle{AllPages}{
\rhead{\thepage}
}
\fancypagestyle{FirstPage}{
\lfoot{\footnotesize{\copyright\space The Authors, 2024 \\ This article is published on arXiv and distributed under the terms of the Creative Commons Attribution Non-Commercial License 4.0 (CC BY-NC 4.0). \\ The definitive, peer reviewed and edited version of this article is published in Power, Energy and Electrical Engineering, Volume 54, Pages 589-600, 2024, doi.org/10.3233/ATDE240359.}}
}

\def\hb{\hbox to 11.5 cm{}}

\begin{document}

\thispagestyle{FirstPage}

\begin{frontmatter}    

\title{A Hybrid Quantum Algorithm for\\Load Flow}

\author[A]{\fnms{David} \snm{Neufeld}},
\author[A]{\fnms{Sajad} \snm{Fathi Hafshejani}
\thanks{Corresponding Author: Sajad Fathi Hafshejani, sajad.fathihafshejan@uleth.ca}},
\author[A]{\fnms{Daya} \snm{Gaur}},
and
\author[A]{\fnms{Robert} \snm{Benkoczi}}

\address[A]{Department of Math and Computer Science, University of Lethbridge, Lethbridge, AB, Canada}

\begin{abstract}
The goal of the load flow study is to ensure that electrical power is delivered efficiently and reliably to end-users while maintaining the stability and security of the power system. Newton-Raphson is a numerical method used widely for load flow analysis. One of the most computationally expensive steps in this method is an equation-solving step. We propose to replace this step with HHL, a quantum algorithm for solving linear systems of equations. HHL is exponentially faster, but with caveats.

In this study, a hybrid quantum algorithm is proposed for solving load flow. The Newton-Raphson method is used as a benchmark to compare the performance of the hybrid quantum algorithm. Although the simulation of the hybrid quantum algorithm takes much time, these preliminary results are encouraging and point to the potential for the use of quantum algorithms to develop hybrid quantum algorithms for load flow analysis and related problems. 
\end{abstract}


\end{frontmatter}

\section{Introduction}

The Load-Flow calculation is a fundamental and widely employed computational tool within the domain of power system analysis. 
The solution to load flow serves as a starting point for various related investigations, including continuation power flow, optimal power flow, and real-time applications \cite{stott1974review,ajjarapu2007computational,kundur2007power}. The analysis of load flow, often referred to as power flow analysis, is a standard procedure in the assessment and design of electrical systems \cite{Ramana2011}. 

The Gauss-Seidel method is one of the earliest techniques employed to address load flow problems \cite{stott1974review,mallick2011development,ebeed2016determination}. However, it is characterized by slow convergence.
Subsequently, the Newton-Raphson (NR) method was introduced to overcome the convergence limitations of Gauss-Seidel \cite{mallick2011development,tinney1967power}. 
Several software for solving load flow exist \cite{Zimmerman_2011, PowerModels, gurobi}. Among the most well-known is MATPOWER \cite{Zimmerman_2011}, a program that can be run in MATLAB. For solving load flow, MATPOWER uses several algorithms, including a Newton-Raphson method \cite{Polyak_2007} based on \cite{Tinney_1967}. Another option for solving load flow is a Julia library called PowerModels.jl \cite{PowerModels}. 
It allows for the use of custom solvers such as Gurobi \cite{gurobi}. These programs for solving load flow take a classical approach. Thus, when solving a nonlinear system of equations, the algorithms require the inversion of the Jacobian matrix. The time required to calculate a matrix inverse using Coppersmith-Winograd type of algorithms is $O(n^{2.373})$ where the matrix is of order $n\times n$. The algorithm due to Strassen \cite{Strassen_1967} has complexity $O(n^{2.807}).$ 

The necessity to compute the inverse of the Jacobian matrix during each iteration to determine the search direction entails a considerable computational burden. 
To tackle this challenge, this paper introduces the use of HHL (Harrow-Hassidim-Lloyd \cite{Harrow_2009}) to determine the search direction. 

{\sc Contributions:} This paper develops a hybrid quantum strategy for solving the load flow problem. Such hybrid quantum algorithms can be used on NISQ quantum computers \cite{saevarsson2022quantum}. We employ the HHL scheme to determine the search direction in each iteration of the Newton-Raphson method. This strategy is exponentially faster in theory \cite{Harrow_2009}. We also conduct a comparative analysis of the performance of classical and quantum methods using simulation studies on two small sized IEEE instances. We solve a 9-bus case; larger than the 5-bus case previously studied \cite{saevarsson2022quantum} albeit on a NISQ quantum computer.

The remainder of this paper is structured as follows: In Section \ref{sec:LF}, we briefly introduce the
load flow problem and Newton-Raphson method.  Section \ref{sec:HHL} discusses the HHL algorithm and the changes made to the Newton-Raphson load flow algorithm.  Section \ref{sec:Res} discusses the setup of our experiment and the results. Finally, in Section \ref{concl}, we summarize our findings and draw conclusions based on our study. We also address the limitations of this approach in the same section.

\section{Related Work}

Several applications for quantum computation in power systems are described in a recent review \cite{golestan2023quantum}. The use of HHL in 
a closely related study is on the development of a quantum approach for the fast decoupled power flow \cite{stott1974fast}. This model uses a constant Jacobian matrix in each iteration, therefore it offers certain advantages, and the authors in \cite{Feng:2021} were able to improve the HHL algorithm when applied to fast decoupled power flow. The use of HHL to develop a hybrid quantum interior point method for solving linear programs was explored in \cite{adoni2023empirical}. It has recently come to our attention that 
AC power flow has been solved on five different NISQ quantum computers for 3-bus and 5-bus cases by \cite{saevarsson2022quantum}. In this paper, we solve a 9-bus case. 
They identified scalability as the main challenge because of the noise. They conclude that as the capability of quantum computers increases, "quantum applications for power systems could become extremely useful for future power system analysis, control, and optimization." Hybrid quantum algorithms for DC power flow have been studied in \cite{eskandarpour2021experimental,eskandarpour2020quantum}.

\section{Load Flow} \label{sec:LF}
This section briefly recalls the load flow problem. Next, we show how to solve the load flow problem by using the Newton-Raphson approach.

\subsection{Load Flow Components} \label{sec:LFC}
The load flow analysis used in balanced and three-phase energy systems under the steady-state condition is generally based on the assumptions described below  \cite{glover2012power}.

\begin{itemize}
     \item Generators meet all load demands connected to the system, and total power loss in transmission lines do not exceed their active and reactive power limits.
    \medskip
    \item Voltage amplitudes of all buses in the system are around the nominal voltage limits.
    \medskip
    \item Transmission lines and transformers are not overloaded.
\end{itemize}

In the load flow analysis, variables to be addressed in each bus include the voltage magnitude $|V|$, the voltage phase angle $\delta$, the active power $P$, and the reactive power $Q$. Some buses are supplied by generators and are called generation buses. In such buses, the voltage amplitude and the active power are assumed to be known (constant). Other buses that are not connected to the generator are called load buses. It is assumed that the complex load power is known in all buses. More details 
can be found in  \cite{kundur2007power}.

Consider a bus $i$ and the lines connected to this bus, from Kirchhoff’s current law, we have
\begin{equation}\label{eq1}
    I_i = \sum_{j=1}^{n} Y_{ij} V_j 
\end{equation}
where $I_i$ denotes the current at bus $i$, $Y_{ij}$ is admittance between bus $i$ and bus $j$, $V_j$ represents voltage (both magnitude and phase angle) at bus $j$, and $n$ is the total number of buses in the system.

The complex power at bus $i$ is given by,
\begin{equation}\label{eq7}
   S_i = V_iI_i^*= V_i \left( \sum_{j=1}^{n} Y_{ij} V_j \right)^* = V_i \sum_{j=1}^{n} Y^*_{ij} V^*_j. 
\end{equation}
Noting that 
\begin{gather}
    Y_{ij} = G_{ij} + j B_{ij}\\
    V_i = |V_i| e^{j\delta_i} = |V_i| \left( \cos \delta_i + j \sin \delta_i \right)
\end{gather}
where $G_{ij}$ and $B_{ij}$ elements are real and complex portions of the admittance function, Eq. (\ref{eq7}) can be written as follows:
\begin{equation}\label{eq8}
    S_i = \sum_{j=1}^{n} |V_i| |V_j| \left( \cos(\delta_i - \delta_j) + j \sin(\delta_i - \delta_j) \right) \left( G_{ij} - j B_{ij} \right).
\end{equation}
This equation can be broken into the active and reactive power components as follows, where $P_i$ ($Q_i$) is the active (reactive) power:
\begin{gather}
    P_i =\sum_{j=1}^{n} |V_i| |V_j| \left( G_{ij} \cos(\delta_k - \delta_j) + B_{ij} \sin(\delta_i - \delta_j) \right)\\
    Q_i = \sum_{j=1}^{n} |V_i| |V_j| \left( G_{ij} \sin(\delta_i - \delta_j) - B_{ij} \cos(\delta_i - \delta_j) \right).
\end{gather}

\subsection{Newton-Raphson for Load Flow} \label{sec:NRLF}
For a detailed description of the Newton-Raphson method 
see \cite{Polyak_2007}. Here we give an overview.
Active and reactive power values of load buses in the power system with $n$ buses can be written as follows \cite{kundur2007power}:
\begin{gather}
  P_i = \sum_{j=1}^{n} Y_{ij} |V_i| |V_j| \cos(\delta_i - \delta_j)\label{eq16} \\
  Q_i = -\sum_{j=1}^{n} Y_{ij} |V_i| |V_j| \sin(\delta_i - \delta_j).\label{eq17}
\end{gather}
Eqs. (\ref{eq16}) and (\ref{eq17}) form the nonlinear equation system containing independent variables (voltage amplitude and voltage phase angle). 
The change in $|V|$ and $\delta$ can be written using 
the Taylor's series for $i \in\{1,2,...,n\}$:
\begin{gather}
 \Delta P_i=\sum_{j=1}^n\frac{\partial P_i}{\partial \delta_j} \Delta\delta_j+\sum_{j=1}^n\frac{\partial P_i}{\partial |V_j|} \Delta |V_j|\label{eq19}\\
 \Delta Q_i=\sum_{j=1}^n\frac{\partial Q_i}{\partial \delta_j} \Delta\delta_j+\sum_{j=1}^n\frac{\partial Q_i}{\partial |V_j|} \Delta |V_j|,\label{eq20}
\end{gather}
where $\Delta P$ and $\Delta Q$ represent power mismatches. They represent the difference between specified powers and calculated powers.
Note that Eqs. (\ref{eq19}) and (\ref{eq20}) can be written in form:
\begin{gather}
 \Delta P_i=\sum_{j=1}^n\frac{\partial P_i}{\partial \delta_j} \Delta\delta_j+\sum_{j=1}^n\frac{\partial P_i}{\partial |V_j|} |V_j|\frac{\Delta |V_j|}{|V_j|}\label{eq21}\\
 \Delta Q_i=\sum_{j=1}^n\frac{\partial Q_i}{\partial \delta_j} \Delta\delta_j+\sum_{j=1}^n\frac{\partial Q_i}{\partial |V_j|} |V_j|\frac{\Delta |V_j|}{|V_j|}.\label{eq22}
\end{gather}
In order to find the solution for the load flow problem, we consider the Jacobian matrix of Eqs. (\ref{eq21}) and (\ref{eq22}) as in \cite{kundur2007power}:
\begin{equation}
    J=\begin{bmatrix}
        \frac{\partial P}{\partial \delta} & \frac{\partial P}{\partial |V|} \\
\frac{\partial Q}{\partial \delta} & \frac{\partial Q}{\partial |V|}
    \end{bmatrix}.
\end{equation}
We define:
\begin{equation}
    \Delta\beta:=
    \begin{bmatrix}
        \Delta P \\
        \Delta Q
    \end{bmatrix}.
\end{equation}
Therefore the direction for Newton's method can be obtained by solving the following system \cite{kundur2007power}:
\begin{equation}\label{search}
   \Delta\alpha:= \begin{bmatrix}
        \Delta\delta \\
        \frac{\Delta V}{\lvert V\rvert}
    \end{bmatrix}
    =
J^{-1}
    \Delta \beta
    ,
\end{equation}
in which $J^{-1}$ denotes the inverse of the matrix Jacobian. 
Now $\delta$ and $|V|$ can be updated  by using the following rules:
\begin{gather}
    \delta^{k+1}=\delta^{k}+\Delta\delta\\
    \lvert V^{k+1}\rvert=\lvert V^{k}\rvert+\frac{\Delta V}{\lvert V^{k}\rvert}\lvert V^{k}\rvert.
\end{gather}

 The Newton-Raphson method is shown in Algorithm \ref{alg:NRLF}. As the number of buses increases, the computation of the inverse of the Jacobian is prohibitively expensive. To address this challenge, we leverage quantum computation techniques. We obtain the search direction using a quantum algorithm (HHL) which does not require explicit computation of the inverse of the Jacobian. Next, we outline the HHL algorithm.

\begin{algorithm}
    \caption{Newton Raphson Load Flow} \label{alg:NRLF}
    \begin{algorithmic}
        \State\textbf{Input:} $P,Q,V,\delta$
        \State Calculate $Y$
        \State Initialize unknown $V/\delta$ values
        \While{Stop condition is not true:}
        \State Compute  $P$ and $Q$ values by using Eqs. (\ref{eq16}) and (\ref{eq17}).
        \State Compute $J$
        \State Solve the system in Eq. (\ref{search})
        \State Update $V$ and $\delta$
        \EndWhile
        \State\textbf{Output:} $V,\delta$
    \end{algorithmic}
\end{algorithm}

\section{HHL} \label{sec:HHL}
HHL is a quantum algorithm for solving linear systems of equations. It is exponentially faster than the fastest known classical algorithm for solving linear systems of equations \cite{Harrow_2009} under some assumptions. In this section, we provide a brief overview of HHL. We also explain how to use HHL to determine the search direction in the Newton-Raphson method. This gives a hybrid quantum algorithm for solving load flow.

\subsection{QFT}
One of the key components of HHL is the Quantum Fourier Transform (QFT). 
QFT is the quantum implementation of the Discrete Fourier Transform (DFT). The QFT is completely specified by its action on each of the basis states, shown below,
\begin{equation}
    \text{QFT} \ket{j} = {\frac{1}{\sqrt{N}}} \sum_{k=0}^{N-1} e^{2\pi i\frac{jk}{N}} \ket{k} 
\end{equation}
where $N=2^n$. The Unitary for QFT can be implemented with only $O(n^2)$  gates which is an exponential improvement compared to $O(N \log{N})$ steps for the classical recursive algorithm. 

\subsection{QPE}
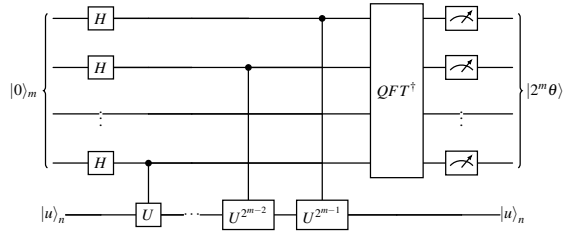
\begin{figure}[htbp]
    \centering
    \begin{tikzpicture} 
    \node[scale=0.6] {
        \begin{quantikz}
            \lstick[4]{\ket{0}$_m$} & \gate{H} & \qw & \qw & \qw & \ctrl{4} & \gate[4]{QFT^\dagger} & \meter{} & \rstick[4]{\ket{2^m\theta}}\\
            & \gate{H} & \qw & \qw & \ctrl{3} & \qw & & \meter{} & \\
            & \vdots & & \qw & \qw & \qw & & \vdots & \\
            & \gate{H} & \ctrl{1} & \qw & \qw & \qw & & \meter{} & \\
            \ket{u}_n & \qw & \gate{U} & \qw\cdots & \gate{U^{2^{m-2}}} & \gate{U^{2^{m-1}}} & \qw & \qw & \ket{u}_n\\
        \end{quantikz}
    };
    \end{tikzpicture}
    \caption{QPE Circuit}
    \label{fig:QPE}
\end{figure}

Another subroutine used within many quantum algorithms, including HHL, is Quantum Phase Estimation (QPE). The input to QPE is a quantum circuit that implements some Unitary $U$. Also given as input is a quantum state $\ket{\psi}$ which is an eigenvector of $U$, i. e.,
\begin{equation}
    U \ket{\psi} = \lambda \ket{\psi}
\end{equation}
The eigenvalue $\lambda$ can be represented as $e^{2\pi i\theta}$. QPE estimates the value $\theta$ up to some precision. 
The circuit that implements QPE uses two quantum registers. The first register has $m$ qubits which specify the precision of $\theta$. The second register contains $n$ qubits used to specify eigenvector $\ket{\psi}$. 
A uniform superposition is created in the first register using the unitary $H^{\otimes m}.$
Next, a controlled version of the unitary $U$ ($C-U$) is applied where the last qubit of the first register is the control. Then, two $C-U$ gates are applied using the second last qubit of the first register as control. This pattern then continues with twice as many gates as previously are applied from the third last qubit of the first register to the second register all the way until $2^{m-1}$ $C-U$ gates are applied using the first qubit of the first register as the control as shown in Figure \ref{fig:QPE}.

An estimate of $\theta$ in the Fourier basis is now in the first register. 
Now, since the estimate of $\theta$ that we have is in the Fourier basis, we need to convert it into the computational basis using the adjoint circuit for QFT, QFT$^\dagger$. 
At this point, the first register contains our estimate of $\theta$ in the computational basis. 
If the value in the first register is $a$, then $\theta=\frac{a}{2^m}$. 
This circuit is shown in Figure \ref{fig:QPE}.

\subsection{HHL Algorithm} \label{sec:HHLalg}
Given the following system of linear equations
\begin{equation}
    Ax=b,
\end{equation}
HHL computes a quantum state $\ket{x}$ which corresponds to the solution $x$.
We assume without loss of generality that $A$ is a Hermitian matrix of size $N\times N$ where $N=2^n$. That is, $A=A^\dagger$ and $x, b$ have appropriate dimensions.

Since $A$ is Hermitian, it can be written in terms of its eigenvalues ($\lambda_j$) and eigenvectors ($\ket{u_j}$) as
\begin{equation}
    A=\sum_{j=0}^{N-1}\lambda_j|u_j\rangle\langle u_j|.
\end{equation}
The inverse is
\begin{equation}
    A^{-1}=\sum_{j=0}^{N-1}\lambda_j^{-1}|u_j\rangle\langle u_j|.
\end{equation}
Let us represent the state $|b\rangle$ in the eigenbasis of $A$ as
\begin{equation}
    |b\rangle=\sum_{j=0}^{N-1}b_j|u_j\rangle.
\end{equation}
The problem of determining $|x\rangle$ can now be written as
\begin{equation}
    |x\rangle=A^{-1}|b\rangle=\sum_{j=0}^{N-1}\lambda_j^{-1}b_j|u_j\rangle.
\end{equation}

The circuit for HHL consists of three registers. The first register contains a single auxiliary qubit which is used for a conditional rotation (more on this later). The second register with $m$ qubits is essentially the first register of QPE. 
Finally, the third register is initially the state 
$|b\rangle$ and contains $|x\rangle$ at the end of the algorithm. 

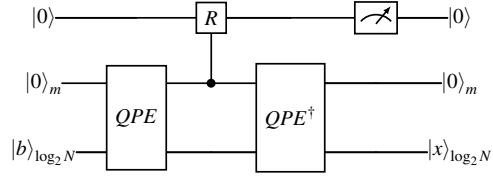
\begin{figure}[htbp]
    \centering
    \begin{tikzpicture}
    \node[scale=0.8] {
        \begin{quantikz}
            \ket{0} & \qw &  \gate[1]{R} & \qw & \meter{} & \qw \ket{0} \\
            \ket{0}_m & \gate[2]{QPE} & \ctrl{-1} & \gate[2]{QPE^\dagger} & \qw & \qw \ket{0}_m \\
            \ket{b}_{\log_2N} & \qw & \qw & \qw & \qw & \qw \ket{x}_{\log_2N}\\
        \end{quantikz}
    };
    \end{tikzpicture}
    \caption{HHL Circuit}
    \label{fig:HHL}
\end{figure}

The first step of HHL is to load $|b\rangle$ into the third register, which now is in state, 
$|b\rangle_{\log_2N}|0\rangle_m|0\rangle.$
We then perform QPE using the second and third registers. 
As $A$ is Hermitian and not Unitary, we use the Unitary $e^{iAt} = \sum_{j=0}^{N-1}e^{i\lambda_jt}|u_j\rangle\langle u_j|,$
where t is some predetermined value. Using QPE we obtain $e^{\lambda_i}$ from which we then obtain $\lambda_i.$
At the end of QPE, the circuit is in the state
\begin{equation}
    \Biggl(\sum_{j=0}^{N-1}b_j|\hat{\lambda}_j\rangle_m|u_j\rangle_{\log_2N}\Biggl)|0\rangle,
\end{equation}
where 
\begin{equation}
    \hat{\lambda}_j=2^m\frac{\lambda_jt}{2\pi}.
\end{equation}
After QPE, we apply a controlled Y-rotation where the second register is the control.  The resulting state is 
\begin{equation}
    \sum_{j=0}^{N-1}b_j|\hat{\lambda}_j\rangle_m|u_j\rangle_{\log_2N}\Biggl(\sqrt{1-\frac{C^2}{\lambda_j^2}}|0\rangle+\frac{C}{\lambda_j}|1\rangle)\Biggl),
\end{equation}
where $C$ is a constant. Next, we apply QPE$^\dagger$. This puts the circuit into the state 
\begin{equation} \label{superposition}
    \sum_{j=0}^{N-1}b_j|0\rangle_m|u_j\rangle_{\log_2N}\Biggl(\sqrt{1-\frac{C^2}{\lambda_j^2}}|0\rangle+\frac{C}{\lambda_j}|1\rangle)\Biggl).
\end{equation}
We then measure the auxiliary qubit. If the result is $|0\rangle$, then we repeat the experiment. If the result is $|1\rangle$, then we have $\lambda_j^{-1}$. At this point, the third register is approximately in the state $|x\rangle$ just as desired. The application of the Unitary results in a superposition governed by Eq. (\ref{superposition}). The probabilities have to be recovered using quantum tomography techniques \cite{vogel1989determination,cramer2010efficient,lanyon2017efficient}. The circuit for HHL is in Figure \ref{fig:HHL}.

\subsection{Newton-Raphson Load Flow With HHL} \label{sec:NRLFHHL}
Solving a linear system of equations can be time-consuming as the number of elements in the vectors and matrices increases. However, it may be possible to improve the time needed to reach a solution by utilizing a quantum algorithm. Thus, we make a modification to the Newton Raphson load flow algorithm discussed in Section \ref{sec:NRLF}. Rather than computing the system in Eq. (\ref{search}) classically, we solve it using HHL.

\begin{algorithm}[htbp]
    \caption{Newton Raphson Load Flow with HHL} \label{alg:NRLFHHL}
    \begin{algorithmic}
        \State\textbf{Input:} $P,Q,V,\delta$
        \State Calculate $Y$
        \State Initialize unknown $V/\delta$ values
        \While{Stop condition is not true:}
        \State Compute new $P$ and $Q$ values
        \State Compute $J$
        \State Solve system in Eq. (\ref{search}) by using HHL
        \State Update $V$ and $\delta$
        \EndWhile
        \State\textbf{Output:} $V,\delta$
    \end{algorithmic}
\end{algorithm}
\section{Experiment and Results} \label{sec:Res}
\subsection{Experiment}

We coded the classical algorithm using Python, and the hybrid quantum algorithm utilized Qiskit 0.42.0 along with the quantum-linear-solvers package for HHL. We ran the algorithm on two widely used test cases: the "Case3" system from \cite{Ramana2011} and "Case9Q" from the MATPOWER cases library \cite{Case9Q}. 

\begin{table}[htbp]
    \centering
    \caption{Case Resource Statistics}
    \begin{tabular}{ |c|c|c|c|c| } \hline
         \textbf{Case} & \textbf{\# of CPUs} & \textbf{Memory} & \textbf{Time Limit} & \textbf{Time Elapsed} \\
         \hline
         Case 3 & 1 & 8 GB & 1 hr & 3 m 12 s \\
         \hline
         Case 9Q & 1 & 64 GB & 72 hr & 12 hr 34 m 19 s \\
         \hline
    \end{tabular}
    \label{table:case_stats}
\end{table}

\subsection{Results}
We note the time required to solve each case using the two different Newton-Raphson load flow algorithms (this information is in Table \ref{table:case_times}). For Case 3, the classical algorithm took 0.0028 seconds, while the hybrid quantum algorithm took 193.9069 seconds. For Case 9Q, the classical algorithm took 0.0169 seconds, while the hybrid quantum algorithm took 40305.7813 seconds. 
The algorithm using HHL took over 69000 times longer to solve Case 3 and over 2384000 times longer to solve Case 9Q because the simulation is exhaustive. This has also been noted in \cite{adoni2023empirical}.
Moreover, the time complexity improvements expected from HHL are highly dependent on the characteristic number of the matrix being used and this requires further study. If indeed the hybrid algorithm was run on a suitable quantum device then each iteration would be faster. The issues of loading the initial state and extracting the solution from the superposition $\ket{x}$ increase the complexity of the hybrid quantum algorithm \cite{adoni2023empirical}.
\begin{table}[htbp]
    \centering
    \caption{Time to Solve Cases}
    \begin{tabular}{ |c|c|c|c|c| } \hline
         \textbf{Case/Version} & \textbf{Time to Solve} \\
         \hline
         Case 3/Classical & 0.0028s \\
         \hline
         Case 3/HHL & 193.9069s \\
         \hline
         Case 9Q/Classical & 0.0169s \\
         \hline
         Case 9Q/HHL & 40305.7813s \\
         \hline
    \end{tabular}
    \label{table:case_times}
\end{table}
We analyzed the iteration counts required by both algorithms. We set the stopping condition as $\Delta\alpha\leq 1e-8$ for both the scenarios. For "Case 3," the classical algorithm converged in 4 iterations, while the hybrid quantum algorithm required 7 iterations.
 For "Case 9Q," the classical algorithm needed 10 iterations to converge, whereas the hybrid quantum algorithm required 11 iterations.

It's important to note that in both cases, the algorithm using HHL required more iterations than the classical algorithm, but it still achieved convergence. Additionally, the increased number of iterations in the HHL-based algorithm did contribute to the overall computational time required in the simulation.

Finally, we look at how significant the difference between the step direction $\Delta\alpha$ obtained by the classical algorithm and the hybrid quantum algorithm is. For this comparison, we compute the second norm of the difference in the two step directions  ($||\Delta\alpha_{CL}-\Delta\alpha_{HHL}||_2$). As seen in Table \ref{table:case3_iter}, the difference between the solutions for Case 3 in the first iteration is less than 0.01 but is still fairly significant. However, in each subsequent iteration, the difference decreases, and when the algorithm with HHL converges in iteration 7, the difference is less than $1e-9$. Similarly, in Case 9Q (shown in Table \ref{table:case9Q_iter}), the difference between the solutions is also less than 0.01 in iteration 1. It then decreases in each subsequent iteration, and when the algorithm with HHL converges in iteration 11, the difference is less than $1e-11$. So, we can see that as more iterations occur, the solution provided by HHL becomes more accurate.
\begin{table}[htbp]
    \centering
    \caption{Iterations of Case 3}
    \begin{tabular}{ |c|c| }
         \hline
         \textbf{Iteration} & $\boldsymbol{||\Delta\alpha_{CL}-\Delta\alpha_{HHL}||_2}$ \\
         \hline
         1 & 0.0036 \\
         \hline
         2 & 0.0003 \\
         \hline
         3 & 2.1845e-05 \\
         \hline
         4 & 1.6773e-06 \\
         \hline
         5 & 1.2872e-07 \\
         \hline
         6 & 9.8784e-09 \\
         \hline
         7 & 7.5807e-10 \\
         \hline
    \end{tabular}
    \label{table:case3_iter}
\end{table}
\begin{table}[htbp]
    \centering
    \caption{Iterations of Case 9Q}
    \begin{tabular}{ |c|c| }
         \hline
         \textbf{Iteration} & $\boldsymbol{||\Delta\alpha_{CL}-\Delta\alpha_{HHL}||_2}$ \\
         \hline
         1 & 0.0034 \\
         \hline
         2 & 0.0005 \\
         \hline
         3 & 1.1144e-05 \\
         \hline
         4 & 1.3193e-07 \\
         \hline
         5 & 3.4950e-08 \\
         \hline
         6 & 8.3729e-09 \\
         \hline
         7 & 1.9885e-09 \\
         \hline
         8 & 4.7059e-10 \\
         \hline
         9 & 1.1138e-10 \\
         \hline
         10 & 2.6360e-11 \\
         \hline
         11 & 6.2386e-12 \\
         \hline
    \end{tabular}
    \label{table:case9Q_iter}
\end{table}

\section{Conclusion}\label{concl}
This paper introduces a hybrid quantum algorithm to solve the load flow problem. We compared the hybrid quantum algorithm with the Newton-Raphson method on standard test cases. The simulation of the hybrid quantum algorithm takes much time. Therefore our experimental study is limited. This preliminary study highlights the potential use of quantum algorithms to develop hybrid quantum algorithms for load flow and related problems which can be solved faster on a quantum machine. Hybrid quantum algorithm hold great promise for speeding up computationally expensive linear algebra operations as identified here. As part of a future study we plan to investigate preconditioning techniques \cite{chen2005matrix} to reduce the condition number of the admittance matrix, and  a modification to the QPE scheme in HHL \cite{Harrow_2009} which is less sensitive to the condition number.

\section*{Acknowledgements}
This research is supported by an NSERC Quantum Alliance Grant and was enabled in part by the Digital Research Alliance of Canada (https://alliancecan.ca/en).

\bibliographystyle{vancouver}
\bibliography{ref}

\end{document}